\documentclass[conference]{IEEEtran}
\usepackage{cite}
\usepackage{amsmath,amssymb,amsfonts}
\usepackage{multirow}
\usepackage{algorithmicx}
\usepackage{graphicx}
\usepackage{textcomp}
\usepackage{xcolor}
\newtheorem{remark}{Remark}  
\def\BibTeX{{\rm B\kern-.05em{\sc i\kern-.025em b}\kern-.08em
    T\kern-.1667em\lower.7ex\hbox{E}\kern-.125emX}}
		
		\makeatletter
\newcommand{\linebreakand}{%
  \end{@IEEEauthorhalign}
  \hfill\mbox{}\par
  \mbox{}\hfill\begin{@IEEEauthorhalign}
}
\makeatother

\begin{document}

\title{First steps toward a simple but efficient model-free control synthesis for variable-speed wind turbines}

\author{\IEEEauthorblockN{Fr\'ed\'eric Lafont}
\IEEEauthorblockA{\textit{Aix Marseille Universit\'e, Universit\'e de Toulon} \\
\textit{CNRS, ENSAM, LIS UMR 7020}\\
13397 Marseille, France \\
lafont@univ-tln.fr}
\and
\IEEEauthorblockN{Jean-Fran\c cois Balmat}
\IEEEauthorblockA{\textit{Aix Marseille Universit\'e, Universit\'e de Toulon} \\
\textit{CNRS, ENSAM, LIS UMR 7020}\\
13397 Marseille, France \\
balmat@univ-tln.fr}
\linebreakand
\IEEEauthorblockN{C\'edric Join}
\IEEEauthorblockA{\textit{Universit\'e de Lorraine},
\textit{CRAN (UMR CNRS 7039)}\\
BP 239, 54506 Vand{\oe}uvre-l\`es-Nancy, France \\
cedric.join@univ-lorraine.fr \\ \& \\
\textit{AL.I.E.N}\\
7 rue Maurice Barr\`es, 54330 V\'ezelise, France \\
cedric.join@alien-sas.com}
\and
\IEEEauthorblockN{Michel Fliess}
\IEEEauthorblockA{\textit{\'Ecole polytechnique},
\textit{LIX (UMR CNRS 7161)}\\
91128 Palaiseau, France \\
Michel.Fliess@polytechnique.edu \\ \& \\
\textit{AL.I.E.N}\\
7 rue Maurice Barr\`es, 54330 V\'ezelise, France \\
michel.fliess@alien-sas.com}
}

\maketitle

\begin{abstract}
Although variable-speed three-blade wind turbines are nowadays quite popular, their control remains a challenging task. We propose a new easily implementable model-free control approach with the corresponding intelligent controllers. Several convincing computer simulations, including some fault accommodations, shows that model-free controllers are more efficient and robust than classic proportional-integral controllers.
\end{abstract}


\section{Introduction}    \label{intro}


Wind energy has been the focus of growing interest for many years. Improving the performances and profitability of wind turbines is currently a burgeoning research topic. Deriving efficient control strategies plays therefore a key r\^{o}le (see, \textit{e.g.}, \cite{b6,b7,brazil,lio,pde}). Advanced techniques are often investigated in the academic literature.

\begin{figure*}[htbp]
\centerline{\includegraphics[width=1.6\columnwidth]{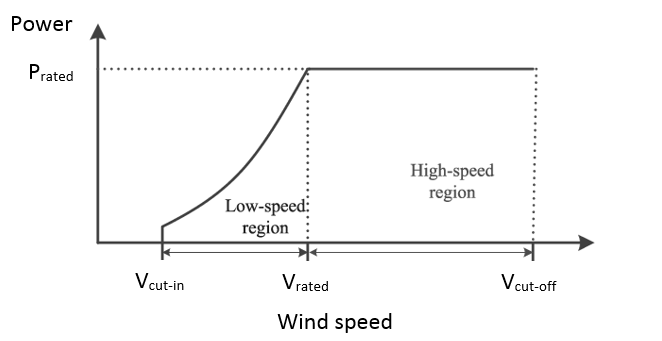}}
\caption{Operating range \cite{b2}}
  \label{Plages}
\end{figure*}

Our study concerns the three-blade machines. They gradually became dominant during the late 1980s. To control such a turbine,  a distinction  (Fig.~\ref{Plages}) is made between the variable speed operating mode (low-speed region) and the power regulation mode (high-speed region) \cite{b1,b2}. In the low-speed region, the wind turbine operates under the nominal power. The purpose of the control is to make maximum use of wind energy and to vary the rotor speed according to the generator torque. In the high-speed region, blade pitch and generator torque can be used for power control. The objective is no longer to maximize the wind energy capture but rather to regulate the energy produced around a nominal value, \textit{i.e.}, around the rated electrical power $P_{\rm rated}$ of the turbine: See Fig.~\ref{Plages}.

Although the severe shortcomings of Proportional-Integral (PI) and Proportional-Integral-Derivative (PID) controllers are well known \cite{astrom,dwyer}, they remain most popular in the industrial world, including wind turbines.  Therefore they are still studied in the academic literature on wind turbines \cite{b3,b4,b5}. 

This paper is devoted to the utilization of \emph{model-free control} (\emph{MFC}) in the sense of \cite{ijc}: 
\begin{itemize}
\item It keeps the benefits of PIs and PIDs, and, especially, the futility of almost any mathematical modeling.
\item Most of the deficiencies of PIs and PIDs are mitigated.
\item Its implementation turns out to be often simpler, when compared to PIs and PIDs.
\end{itemize}
MFC has been successfully applied all over the world as demonstrated by the references in \cite{ijc,bara,ftc}. See already \cite{isa,flateol} for its relevance to wind turbine. We show here the interest of MFC to control the pitch angle and the generator torque for low and high winds.
\begin{remark}
The enormous difficulty of writing down a suitable mathematical modeling of wind turbines explains why other model-free settings have been proposed (see, \textit{e.g.}, \cite{iet}). They are mostly based on various optimization techniques.
\end{remark}
\begin{remark}
A patent (\emph{\'Electricit\'e de France} (\emph{EDF}) and \emph{\'Ecole polytechnique}) has been associated to the use of MFC  for hydroelectric power plants \cite{cifa}, \textit{i.e.}, to another green energy production.
\end{remark}

Our paper is organized as follows. For the sake of computer simulations Section \ref{syst} provide short mathematical descriptions of wind turbines and of the wind. MFC is briefly recalled in Section \ref{free}. Computer simulations are presented in Section \ref{Res} where fault accommodations are considered: They show a striking superiority of MFC with respect to a classical PI. Some concluding suggestions for future research are sketched in Section \ref{conclu}.


\section{Mathematical modelings for \\ computer simulations} \label{syst}
\subsection{Wind turbine}
Let us emphasize that the model below is more or less improper for deriving control laws. What follows is borrowed from \cite{b2,b10,b11,boukh}.

See Table \ref{Param} for some useful parameters where

\begin{table}[htbp]
\caption{Wind turbine parameters}
\label{Param}
\begin{center}
\begin{tabular}{|c||c|}
\hline
Variable & Value \\
\hline
$J_{\rm t}$ ($kgm^2$) & $3.89 \cdot 10^{5}$ \\
\hline
$K_{\rm t}$ ($Nm/rad/s$)  & 400 \\
\hline
$\rho$ ($kgm^3$) & 1.29 \\
\hline
R ($m$) & 21.65 \\
\hline
$max \ T_{g}$ ($kNm$) & 162 \\
\hline
$P_{\rm rated}$ ($kW$) & 600 \\
\hline
\end{tabular} 
\end{center}
\end{table}

\begin{itemize}
\item $\rho$ is the air density (kg/m$^3$),
\item $R$ is the radius of the blade (m),
\item $V$ is the wind speed (m/s),
\item $T_{\text{g}}$ is the generator torque (Nm), 
\item ${\rm max} \ T_{\rm g}$ is the maximum value of $T_{\text{g}}$,
\item $J_{\rm t}$ is the combined inertia of the turbine and generator (kgm$^2$),
\item $K_{\rm t}$  is the the damping coefficient of the turbine (Nm/rad/s),
\item $C_p\left(\lambda,\beta\right)$ is the power conversion coefficient. It depends on the tip-speed ratio $\lambda$ and the pitch angle $\beta$:


	\begin{equation*}
		C_p\left(\lambda,\beta\right)=c_1\left(\frac{c_{2}}{\lambda_{i}}-c_3 \beta - c_4\right) \exp\left(\frac{-c_{5}}{\lambda_{i}}\right) + c_6 \lambda
		\label{equa3}
		\end{equation*}
		
		\begin{equation*}
		\frac{1}{\lambda_{i}}=\frac{1}{\lambda+0.08 \beta} -  \frac{0.035}{\beta^3+1}
		\label{equa4}
		\end{equation*}
See Table \ref{Paramcoeff} for the coefficients $c_\kappa$, $\kappa = 1, \dots, 6$.

\begin{table}[htbp]
\caption{Coefficients for the power conversion coefficient}
\label{Paramcoeff}
\begin{center}
\begin{tabular}{|c||c|}
\hline
Coefficient & Value \\
\hline
$c_{\rm 1}$ & 0.4 \\
\hline
$c_{\rm 2}$ & 116 \\
\hline
$c_{\rm 3}$ & 0.4 \\
\hline
$c_{\rm 4}$ & 5 \\
\hline
$c_{\rm 5}$ & 21 \\
\hline
$c_{\rm 6}$ & 0.02 \\
\hline
\end{tabular} 
\end{center}
\end{table}
\end{itemize}
The output power of wind turbines is given by
\begin{equation*}
		P_{t}\  (W)  = \frac{1}{2} \rho \pi R^2  C_p\left(\lambda,\beta\right) V^{3}
		\label{equa1}
\end{equation*}

\noindent The tip-speed ratio is defined by 

\begin{equation*}
		\lambda = \frac{R \omega_{t}}{V}
		\label{equa2}
		\end{equation*}
where $\omega_{t}$ is the turbine angular speed (rad/s).

\begin{figure*}[htbp]
\centerline{\includegraphics[width=1.6\columnwidth]{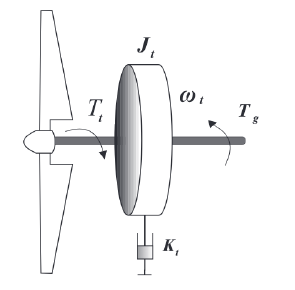}}
\caption{One-mass model of a wind turbine \cite{boukh}}
  \label{eolescheme}
\end{figure*}

\noindent Write the turbine torque
	
		\begin{equation*}
		T_{\rm t} = \frac{1}{2} \rho \pi R^3 V^2 \frac{C_p\left(\lambda,\beta\right)}{\lambda}
		\label{equa8}
		\end{equation*}
If a perfectly rigid low-speed shaft is assumed, a one-mass model of the turbine (see Fig.~\ref{eolescheme}) may be expressed by
     
  \begin{equation*}
	J_{\rm t} \dot{\omega}_{\rm t}	= T_{\rm t} - K_{\rm t} \omega_{\rm t} - T_{\text{g}}
			\label{equa9}
	\end{equation*}
	


\subsection{Wind}
The wind speed $V$ may vary considerably. Its variation may be modeled as a finite sum of harmonics in the frequency range 0.1-10 Hz:

		\begin{equation}\label{wind} 
		V = V_{\text{moy}} \left[1+ \sum_{\rm finite} A_k \sin(\omega_k t)\right]
		\end{equation}
$V_{\text{moy}}$ is the mean speed, $A_k$ is the magnitude of the $k$th sine wave of frequency $\omega_k$ \cite{b20}. Eq. \eqref{wind} reads here

		\begin{multline*} \label{modelV}
		V = V_{\text{moy}} \left[1+ A_1 \sin(0.1047t) + A_2 \sin(0.2674t)\right.\\
		\left.+ A_3 \sin(1.309t) + A_4 \sin(3.696t)\right]
		\end{multline*}
See Table \ref{coefficients} for the coefficients $A_\iota$, $\iota = 1, \dots, 4$.

\begin{table}[htbp]
\caption{Coefficient magnitudes with respect to the mean wind speed}
\label{coefficients}
\begin{center}
{\scriptsize
\begin{tabular}{|c||c|c|c|c|}
\hline
Mean wind speed & \multicolumn{4}{|c|}{Coefficients} \\
\hline
$V_{\text{moy}}$ (m/s) & $A_1$ & $A_2$ & $A_3$ & $A_4$ \\
\hline
 7 & 0.029 & 0.286 & 0.143 & 0.029 \\
\hline
 8 & 0.025 & 0.25 & 0.125 & 0.025 \\
\hline
 9 & 0.022 & 0.222 & 0.111 & 0.022 \\
\hline
 16 & 0.0125 & 0.125 & 0.0625 & 0.0125 \\
\hline
 20 & 0.01 & 0.1 & 0.05 & 0.01 \\
\hline
\end{tabular}}
\end{center}
\end{table}

\section{Model-free control \\ and intelligent controllers\protect\footnote{See \cite{ijc} for more details.}} \label{free}

For the sake of notational simplicity, let us restrict ourselves to single-input single-output (SISO) systems. 

\subsection{The ultra-local model}
The unknown global description of the plant is replaced by the following first-order \emph{ultra-local model}:
\begin{equation}
\boxed{\dot{y} = F + \alpha u} \label{1}
\end{equation}
where:
\begin{enumerate}
\item The control and output variables are respectively $u$ and $y$.
\item $\alpha \in \mathbb{R}$ is chosen by the practitioner such that the three terms in Equation \eqref{1} have the same magnitude.
\end{enumerate}
The following comments are useful:
\begin{itemize}
\item $F$ is \emph{data driven}: it is given by the past values of $u$ and $y$.
\item $F$ includes not only the unknown structure of the system but also any disturbance.
\end{itemize}




\subsection{Intelligent controllers}
Close the loop with the \emph{intelligent proportional controller}, or \emph{iP}, 
\begin{equation}\label{ip}
\boxed{u = - \frac{F_{\text{est}} - \dot{y}^\ast + K_P e}{\alpha}}
\end{equation}
where
\begin{itemize}
\item $y^\ast$ is the reference trajectory,
\item $e = y - y^\star$ is the tracking error,
\item $F_{\text{est}}$ is an estimated value of $F$
\item $K_P \in \mathbb{R}$ is a gain.
\end{itemize}
Equations \eqref{1} and \eqref{ip} yield
\begin{equation*}
	\dot{e} + K_P e = F - F_{\text{est}}
			\label{equa15}
	\end{equation*}
If the estimation $F_{\text{est}}$ is ``good'': $F - F_{\text{est}}$ is ``small'', \textit{i.e.}, $F - F_{\text{est}} \simeq 0$,  then $\lim_{t \to +\infty} e(t) \simeq 0$ if $K_P > 0$. It implies that the tuning of $K_P$ is straightforward. This is a major benefit when
compared to the tuning of ``classic'' PIDs (see, \textit{e.g.}, \cite{astrom,dwyer,franklin}).

\subsection{Estimation of $F$}\label{F}
Mathematical analysis \cite{b15} tells us that under a very weak integrability assumption, any function, for instance $F$ in Eq. \eqref{1}, is ``well'' approximated by a piecewise constant function.
\subsubsection{First  approach}
Rewrite then Eq. \eqref{1}  in the operational domain \cite{b18}: 

\begin{equation*}
sY = \frac{\Phi}{s}+\alpha U +y(0)
			\label{equa16}
	\end{equation*}
where $\Phi$ is a constant. We get rid of the initial condition $y(0)$ by multiplying both sides on the left by $\frac{d}{ds}$:

\begin{equation*}
Y + s\frac{dY}{ds}=-\frac{\Phi}{s^2}+\alpha \frac{dU}{ds}
			\label{equa17}
	\end{equation*}
	
Noise attenuation is achieved by multiplying both sides on the left by $s^{-2}$, \textit{i.e.}, via integration \cite{noise}. It yields in the time domain the real-time estimate, thanks to the equivalence between $\frac{d}{ds}$ and the multiplication by $-t$,
\begin{equation*}\label{integral}
{\small F_{\text{est}}(t)  =-\frac{6}{\tau^3}\int_{t-\tau}^t \left\lbrack (\tau -2\sigma)y(\sigma)+\alpha\sigma(\tau -\sigma)u(\sigma) \right\rbrack d\sigma }
\end{equation*}
where $\tau > 0$ might be quite small. This integral, which is a low pass filter, may of course be replaced in practice by a classic digital filter.

\subsubsection{Second approach}\label{2e}
Close the loop with the iP \eqref{ip}. It yields:

\begin{equation*}
F_{\text{est}}(t) = \frac{1}{\tau}\left[\int_{t - \tau}^{t}\left(\dot{y}^{\star}-\alpha u
- K_P e \right) d\sigma \right] 
			\label{equa18}
	\end{equation*}
\begin{remark}	
Let us emphasize that implementing our intelligent controllers is easy \cite{ijc}, \cite{b19}.
\end{remark}

\subsection{MIMO systems}
Consider a multi-input multi-output (MIMO) system with $m$ control variables $u_i$ and $m$ output variables $y_i$, $i = 1, \dots, m$, $m \geq 2$. It has been observed in \cite{toulon} and confirmed by all encountered concrete case-studies (see, \textit{e.g.}, \cite{wang}), that such a system may usually be regulated via $m$ monovariable ultra-local models:
\begin{equation*}\label{multi}
y_{i}^{(n_i)} = F_i + \alpha_i u_i
\end{equation*}
where $F_i$ may also depend on $u_j$, $y_j$, and their derivatives, $j \neq i$.

\section{Simulation results} \label{Res}
Write $P_{\rm e}$ the electrical power of the turbine. Define, according to Fig.~\ref{Plages}, the following regions:
\begin{itemize}
\item Low-speed region:  $V_{\rm cut-in} \leq V < V_{\rm rated}$, $P_{\text{e}}<P_{\rm rated}$.
\item High-speed region: $V_{\rm rated} \leq V \leq V_{\rm cut-off}$, $P_{\text{e}}=P_{\rm rated}$.
\end{itemize}



\subsection{Simulation with an alternating wind in the low-speed region} \label{Addres}
\begin{figure*}[htbp]
\centerline{\includegraphics[width=1.6\columnwidth]{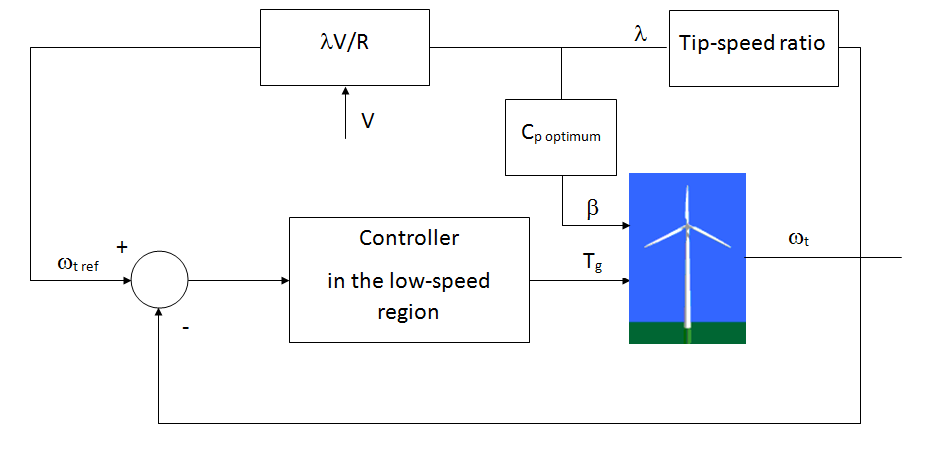}}
\caption{Block diagram in the low-speed region}
  \label{contrvv}
\end{figure*}

\begin{figure*}[htbp]
\centerline{\includegraphics[width=1.6\columnwidth]{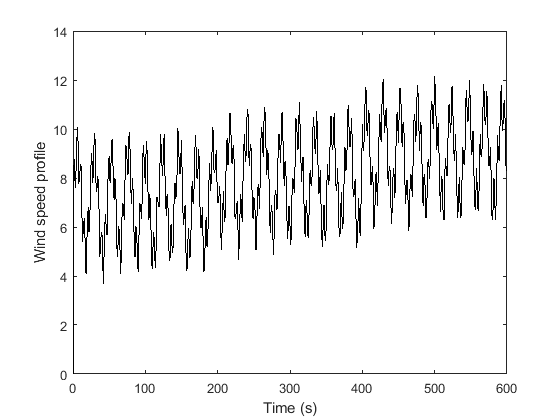}}
\caption{Alternating wind in the low-speed region}
      \label{Vvmixfaible}
\end{figure*}

Fig.~\ref{contrvv} presents the monovariable structure of the controller with a low-speed wind. See
\begin{itemize}
\item if $t<200$ s then $V_{\text{moy}} = 7$ m/s,
\item if $t>200$ s, $t<400$ s then $V_{\text{moy}} = 8$ m/s,
\item if $t>400$ s then $V_{\text{moy}} = 9$ m/s,
\end{itemize} 
and Fig.~\ref{Vvmixfaible}. Performances of PI and iP controllers are compared. 



\subsubsection{iP} \label{ResVfaible}
The parameters  $K_P$ and $\alpha$ are given in the Table \ref{Contr1}. As explained in Section \ref{free}, $\tau$ must be chosen quite small.

\begin{figure*}[htbp]
\centerline{\includegraphics[width=1.6\columnwidth]{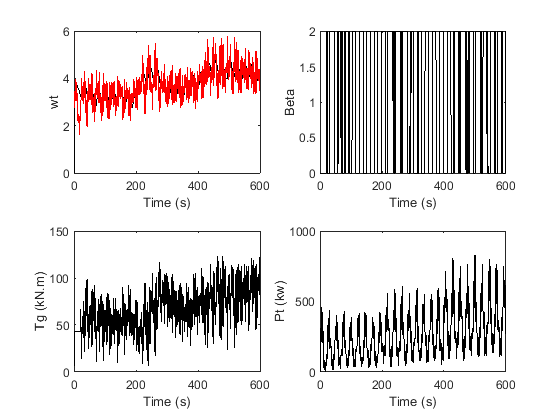}}
\caption{Control of $\omega_{\rm t}$ and $P_e$ with an alternating wind in the low-speed region (iP controller)}
 \label{resulmixfaible}
\end{figure*}

\begin{table}[htbp]
\caption{iP controller parameters for the weak wind}
\label{Contr1}
\begin{center}
\begin{tabular}{|c||c|}
\hline
Variable & Control of $T_{\text{g}}$ \\
\hline
$K_P$ & -0.45 \\
\hline
$\alpha$  & 0.0005 \\
\hline
$\tau$  & 20 \\
\hline
\end{tabular} 
\end{center}
\end{table}
\noindent Fig.~\ref{resulmixfaible} shows the efficiency of the iP controller:
\begin{itemize}
\item The turbine angular speed $\omega_{\rm t}$ oscillates around the reference value. 
\item This reference value varies in function of the wind speed value. 
\item The blade pitch angle $\beta$ varies between 0 and 2 degrees.  
\item Blades are positioned to recover the maximum energy.
\item The generator torque $T_{\text{g}}$ is limited to $162$ kNm. 
\end{itemize}
		

\subsubsection{PI} \label{ResVfaiblePID}
\begin{figure*}[htbp]
\centerline{\includegraphics[width=1.6\columnwidth]{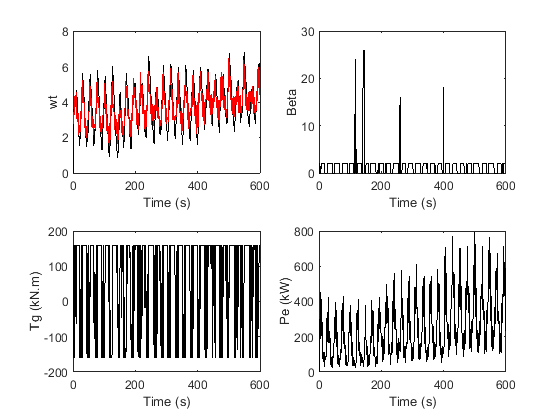}}
\caption{Control of $\omega_{\rm t}$ and $P_e$ with alternating wind in the low-speed region (PI controller)}
      \label{resulmixfaiblePID}
\end{figure*}

For $T_{\text{g}}$, the Ziegler-Nichols method (see, \textit{e.g.}, \cite{franklin}) yield the coefficients of the PI controller: seeTable \ref{Contr1PI}:

\begin{table}[htbp]
\caption{PI controller parameters for the weak wind}
\label{Contr1PI}
\begin{center}
\begin{tabular}{|c||c|}
\hline
Variable & Control of $T_{\text{g}}$ \\
\hline
$K_P$ & {500} \\
\hline
$K_I$ & {10} \\
\hline
\end{tabular} 
\end{center}
\end{table}
\noindent See Fig.~\ref{resulmixfaiblePID} for the results.


\subsubsection{Performances comparison of the two controllers} \label{Comp1}
The aim is to maximize the wind energy $P_{\rm t}$. The mean absolute error (MAE) and the corresponding standard deviation for each controller are evaluated for the iP and PI: See Table \ref{compa3}. The mean of the output power, which is calculated between $t = 60$s, where the steady state is established, and the final time $t = 600$s, is higher with the iP. The iP is thus more efficient. 
		
\begin{table}[htbp]
\caption{MAE and standard deviation comparison for $\omega_{\rm t}$ and $P_{\rm t}$ with an alternating wind in the low-speed region}
\label{compa3}
\begin{center}
{\scriptsize
\begin{tabular}{|c|c|c||c|c|}
\hline
\multicolumn{3}{|c||}{Controller} & iP & PI \\
\hline
\multirow{3}*{Weak wind} & \multirow{2}*{$\omega_{\rm t}$ (rad/s)} & MAE & {0.43} & {0.61} \\
\cline{3-5}
 & & standard deviation & {0.55} & {0.70} \\
\cline{2-5}
 & $P{\rm t}$ (kW) & MAE & {279} & {278} \\
\hline
\end{tabular}}
\end{center}
\end{table}

\subsection{Simulation with an alternating wind in the high-speed region}
\begin{figure*}[htbp]
\centerline{\includegraphics[width=1.6\columnwidth]{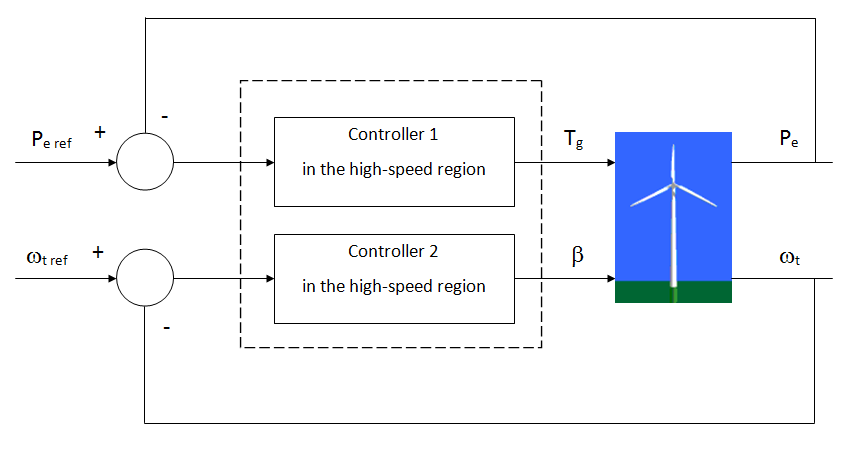}}
\caption{Block diagram in the high-speed region}
  \label{contrvf}
\end{figure*}

\begin{figure*}[htbp]
\centerline{\includegraphics[width=1.6\columnwidth]{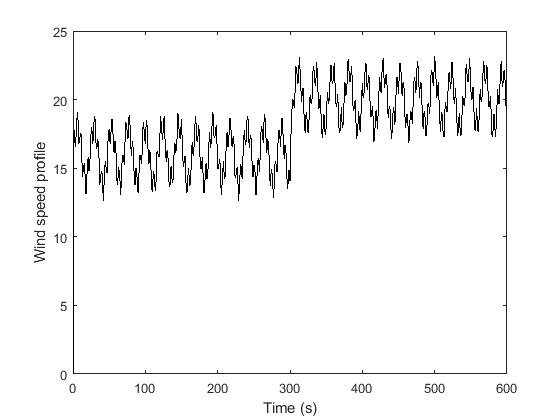}}
\caption{Alternating wind in the high-speed region}
      \label{Vvmixfort}
\end{figure*}

Fig.~\ref{contrvf} exhibits the new MIMO control structure with $2$ control and output variables. See Fig.~\ref{Vvmixfort} for the wind. There are two operating points:
\begin{itemize}
\item if $t<300$ s then $V_{\text{moy}} = 16\;$ m/s,
\item if $t>300$ s then $V_{\text{moy}} = 20\;$ m/s.
\end{itemize}


\subsubsection{iPs} \label{ResVfort}
\begin{figure*}[htbp]
\centerline{\includegraphics[width=1.6\columnwidth]{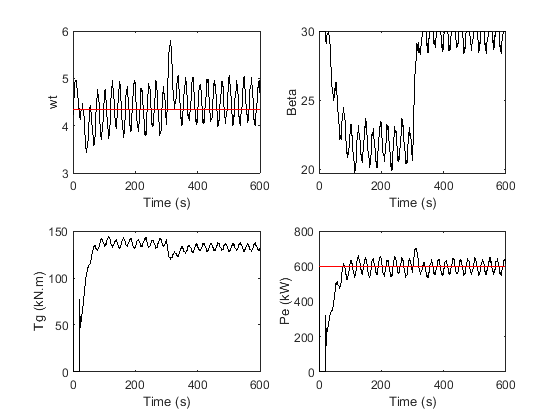}}
\caption{Control of $\omega_{\rm t}$ and $P_e$ with alternating wind in the high-speed region (iP controller)}
      \label{resulmixfort}
\end{figure*}

See Table \ref{Contr2} for the choice $K_P$, $\alpha$ and $\tau$. As depicted by Fig.~\ref{resulmixfort}, 
\begin{itemize}
\item $P_{\rm e}$ remains stable around $P_{\rm rated}$ = 600 kW,
\item $\omega_{\rm t}$ remains close to the reference, 
\item \begin{enumerate}
\item the initial value of $\beta$ is equal to 30 degrees, 
\item then it is approximately equal to 21 degrees until $t = 300$s when the steady state is established,
\item for $t > 300$s it fluctuates around 29 degrees when the mean wind speed increases to 20 m/s,
\end{enumerate}
\item the value of $T_g$ stays around 130 kNm when the steady state is established.
\end{itemize}

\begin{table}[htbp]
\caption{Model-free control parameters for the strong wind}
\label{Contr2}
\begin{center}
\begin{tabular}{|c||c|c|}
\hline
Variable & Control of $\beta$ &  Control of $T_{\text{g}}$ \\
\hline
$K_P$ & -4 & 3 \\
\hline
$\alpha$  & 1 & 1000 \\
\hline
$\tau$  & 20  & 20 \\
\hline
\end{tabular} 
\end{center}
\end{table}
	
	
\subsubsection{PIs} \label{ResVfortPID}
\begin{figure*}[htbp]
\centerline{\includegraphics[width=1.6\columnwidth]{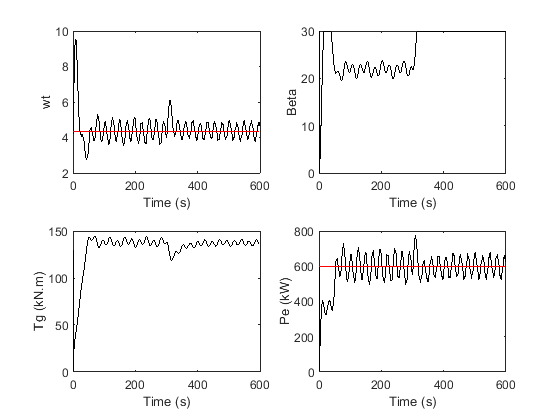}}
\caption{Control of $\omega_{\rm t}$ and $P_e$ with alternating wind in the high-speed region (PI controller)}
 \label{resulmixfortPID}
\end{figure*}

See Table \ref{Contr2PI} for the gains of the two PIs which are again determined via the Ziegler-Nichols method. According to Fig.~\ref{resulmixfortPID}, their performances are poorer.

\begin{table}[htbp]
\caption{PI controller parameters for the strong wind}
\label{Contr2PI}
\begin{center}
\begin{tabular}{|c||c|c|}
\hline
Variable & Control of $\beta$ &  Control of $T_{\text{g}}$ \\
\hline
$K_P$ & -0.006 & -0.0003 \\
\hline
$K_I$ & 0.52 & -0.00026 \\
\hline
\end{tabular} 
\end{center}
\end{table}

	

\subsubsection{Performances comparison of the two controllers} \label{Comp2}
The main objective is not to maximize the electrical production but to maintain it close to $P_{\rm rated}$ = 600 kW as shown in Fig.~\ref{Plages}. As in Section \ref{Comp1} MAE and standard deviation are reported  in Table \ref{compa4}. They confirm the marked superiority of iPs.

\begin{remark}
The advanges of intelligent controllers with respect to PIDs were already noticed several times. See \cite{madrid,mercer} for two recent most convincing examples.
\end{remark}
	
\begin{table}[htbp]
\caption{MAE and standard deviation comparison for $\omega_{\rm t}$ and $P_e$ with an alternating wind in the high-speed region}
\label{compa4}
\begin{center}
{\scriptsize
\begin{tabular}{|c|c|c||c|c|}
\hline
\multicolumn{3}{|c||}{Controller} & Intelligent P & Classical PI \\
\hline
\multirow{4}*{Strong wind} &  \multirow{2}*{$\omega_{\rm t}$ (rad/s)} & mean & 0.33 & 0.38 \\
\cline{3-5}
 & & standard deviation & 0.39 & 0.45 \\
\cline{2-5}
 &  \multirow{2}*{$P_e$ (kW)} & mean & 32 & 48 \\
\cline{3-5}
 & & standard deviation & 38 & 56 \\
\hline
\end{tabular}}
\end{center}
\end{table}

\subsection{Fault accommodation and iP}
Simulate, as proposed by \cite{b21}, an actuator fault, occurring at $t = 300$ s with a wind speed equal to $16$ m/s, with respect to the electromagnetic torque $T_{\text{g}}$:
\begin{itemize}
\item a loss of efficiency equal to 15\%, \textit{i.e.},
	\begin{equation*}
		T_{{\text{g\  actuator}}}  = 0.85 \times T_{\rm g}
		\label{effi}
	\end{equation*}
where $T_{\text{g\  actuator}}$ is the control applied to the turbine;

\item a bias fault equal to 50 kNm, \textit{i.e.},
\begin{equation*}
		T_{\text{g\  actuator}}  = T_{\rm g} -50
		\label{effi}
	\end{equation*}
\end{itemize}

\subsubsection{Loss of efficiency}\label{effi}
\begin{figure*}[htbp]
\centerline{\includegraphics[width=1.6\columnwidth]{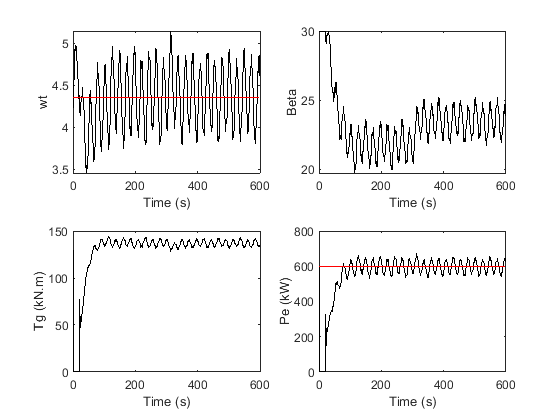}}
\caption{Fault accomodation with iP (loss of efficiency)}
      \label{resulfaibledefault1}
\end{figure*}

Fig.~\ref{resulfaibledefault1} depicts the good results for the iP. Let us emphasize that our setting is modifying $\beta$.
	
	
\subsubsection{Bias fault}
\begin{figure*}[htbp]
\centerline{\includegraphics[width=1.6\columnwidth]{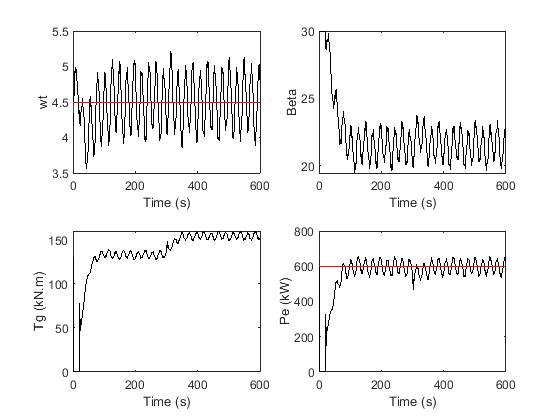}}
\caption{Fault accommodation with iP (bias fault)}
      \label{resulfaibledefault2}
\end{figure*}

Fig.~\ref{resulfaibledefault2} exhibits similar results to those in Section \ref{effi}. 
	

\section{Conclusion} \label{conclu}
The promising simulations depicted here need of course to be confirmed, especially via some concrete plants. The behavior of our control strategy with respect to the unavoidable vibrations (see, \textit{e.g.}, \cite{vibration1,vibration2}) should of course be clarified. For a better energy management, some forecasting of the wind power is crucial. It would be rewarding to extend the time series techniques for photovoltaic  energy \cite{solar}, which bears some similarity with MFC. 

\newpage

\end{document}